\documentclass[11pt]{article}
\usepackage{amsmath,amsthm,amsfonts}
\usepackage{amssymb,graphics}  

\newtheorem{theorem}{Theorem}

\newcommand{\be}{\begin{eqnarray}}
\newcommand{\ee}{\end{eqnarray}}

\begin{document}

\title{\LARGE\bf From the Kadomtsev-Petviashvili equation \\
  halfway to Ward's chiral model 
}

\date{}

\author{\Large Aristophanes Dimakis$^a$ and Folkert M\"uller-Hoissen$^b$ \\ \\
 \normalsize
 $^a$ Department of Financial and Management Engineering, 
 University of the Aegean \\
 31 Fostini Str., GR-82100 Chios, Greece \\
 $^b$ Max-Planck-Institute for Dynamics and Self-Organization \\
 Bunsenstrasse 10, D-37073 G\"ottingen, Germany \\ 
 E-mails: dimakis@aegean.gr, folkert.mueller-hoissen@ds.mpg.de }

\maketitle

\thispagestyle{empty}

\begin{abstract}
The ``pseudodual'' of Ward's modified chiral model is a dispersionless limit 
of the matrix Kadomtsev-Petviashvili (KP) equation. 
This relation allows to carry solution techniques from KP over to the former model. 
In particular, lump solutions of the $su(m)$ model with rather 
complex interaction patterns are reached in this way. 
We present a new example. 
\end{abstract}

\setcounter{equation}{0}
Ward's chiral model in $2+1$ dimensions \cite{Ward88} (see \cite{DMH07disp} 
for further references) is given by 
\be
  (J^{-1} J_t)_t - (J^{-1} J_x)_x - (J^{-1} J_y)_y 
      + [J^{-1} J_x , J^{-1} J_t] = 0   \label{Ward}
\ee
for an $SU(m)$ matrix $J$, where $J_t = \partial J/\partial t$, etc. 
In terms of the new variables  
\be
    x_1 := (t-x)/2 \, , \quad x_2 := y \, , \quad x_3 := (t+x)/2 \, ,  
               \label{coord_transf}
\ee
this simplifies to 
$(J^{-1} J_{x_3})_{x_1} - (J^{-1} J_{x_2})_{x_2} = 0$, 
which extends to the hierarchy
\be
   (J^{-1} J_{x_{n+1}})_{x_m} - (J^{-1} J_{x_{m+1}})_{x_n} = 0 \, , 
   \qquad m,n = 1,2,  \ldots \; .   \label{Ward_hier}
\ee
The Ward equation is completely integrable\footnote{In the sense of the 
inverse scattering method, the existence of a hierarchy, and various other 
characterisations of complete integrability. In the following ``integrable'' 
loosely refers to any of them. }  
and admits soliton-like solutions, often called ``lumps''. 
It was shown numerically \cite{Sutc92} and later analytically \cite{Ward95,Ioan96,Dai+Tern07} 
that such lumps can interact in a nontrivial way, unlike usual solitons.
In particular, they can scatter at right angles, a phenomenon sometimes referred 
to as ``anomalous scattering''.\footnote{See also the references cited above 
for related work. Anomalous scattering has also been found in some related 
non-integrable systems, like sigma models, Yang-Mills-Higgs 
equation (monopoles) and the Abelian Higgs model or Ginzburg-Landau equation 
(vortices), see \cite{Mant+Sutc04} for instance. } 
Also the integrable KP equation, more precisely KP-I (``positive dispersion''), possesses 
lump solutions with anomalous scattering \cite{GPS93JETP,Vill+Ablo99,ACTV00} 
(besides those with trivial scattering \cite{MZBIM77}). 
Introducing a potential $\phi$ for the real scalar function $u$ via $u = \phi_x$, 
in terms of independent variables $t_1,t_2$ (spatial coordinates) and $t_3$ (time), 
the (potential) KP equation is given by 
\be  
  ( 4 \, \phi_{t_3} - \phi_{t_1t_1t_1}
    - 6 \, \phi_{t_1} \, \phi_{t_1} )_{t_1} - 3 \sigma^2 \, \phi_{t_2t_2}
  = 0 \, ,
\ee
with $\sigma=i$ in case of KP-I and $\sigma=1$ for KP-II. Could it be that this equation 
has a closer relation with the Ward equation? We are trying to compare an equation 
for a scalar with a matrix equation, and in \cite{Ward95} the appearance of 
nontrivial lump interactions in the Ward model had been attributed to the presence 
of the ``internal degrees of freedom'' of the latter. At first sight this does 
not match at all. 
However, the resolution lies in the fact that the KP equation possesses an 
integrable extension to a (complex) matrix version, 
\be  
  \Big( 4 \, \Phi_{t_3} - \Phi_{t_1t_1t_1}
    - 6 \, \Phi_{t_1} \, Q \, \Phi_{t_1} \Big)_{t_1} - 3 \sigma^2 \, \Phi_{t_2t_2}
  = - 6 \sigma \, [\Phi_{t_1} , \Phi_{t_2}]_Q  \, ,
\ee
where we modified the product by introducing a constant $N \times M$ matrix $Q$,  
and the commutator is modified accordingly, so that $[\Phi_{t_1} , \Phi_{t_2}]_Q 
 = \Phi_{t_1} Q \Phi_{t_2} - \Phi_{t_2} Q \Phi_{t_1}$. 
Here $\Phi$ is an $M \times N$ matrix. 
If $\mathrm{rank}(Q)=1$, and thus $Q = V U^\dagger$ with vectors $U$ and $V$, 
then any solution of this (potential) matrix KP equation determines a solution 
$\phi := U^\dagger \Phi V$ of the scalar KP equation.\footnote{See 
e.g. \cite{March88,Carl+Schi99} for related ideas. }
More generally, this extends to the corresponding (potential) KP hierarchies. 

Next we look for a relation between the matrix KP and the Ward equation. 
Indeed, there is a dispersionless (multiscaling) 
limit of the above ``noncommutative'' (i.e. matrix) KP equation, 
\be
  \Phi_{x_1 x_3} - \sigma^2 \, \Phi_{x_2 x_2} = - \sigma \, [\Phi_{x_1} , \Phi_{x_2} ]_Q \, , 
          \label{QpdCM}
\ee
obtained by introducing $x_n = n \, \epsilon \, t_n$ with a parameter $\epsilon$, 
and letting $\epsilon \to 0$ (assuming an appropriate dependence of the KP variable $\Phi$ 
on $\epsilon$) \cite{DMH07disp}. If  $\mathrm{rank}(Q)=m$, and thus $Q = V U^\dagger$ 
with an $M \times m$ matrix $U$ and an $N \times m$ matrix $V$, then the 
$m \times m$ matrix $\varphi := \sigma \, U^\dagger \Phi V$ solves 
\be
  \varphi_{x_1 x_3} - \sigma^2 \, \varphi_{x_2 x_2} = - [\varphi_{x_1} , \varphi_{x_2} ] \, , 
             \label{pdCM}
\ee
if $\Phi$ solves (\ref{QpdCM}). In terms of the variables $x,y,t$, this 
becomes\footnote{This Leznov equation \cite{Lezn87} and the Ward equation arise by gauge-fixing of the 
hyperbolic Bogomolny equation, see e.g. \cite{Duna+Mant05}. }
\be
    \varphi_{tt} - \varphi_{xx} - \sigma^2 \, \varphi_{yy} 
          + [ \varphi_t - \varphi_x , \varphi_y ] =0 \; .  \label{pdCMv2}
\ee
Now we note that the cases $\sigma=i$ and $\sigma=1$ are related by exchanging 
$x$ and $t$, hence they are equivalent.\footnote{Note also that this transformation 
leaves the conserved density (\ref{E}) invariant.} 
We choose $\sigma=1$ in the following. Then (\ref{pdCM}) extends to the hierarchy
\be
  \varphi_{x_m x_{n+1}} - \varphi_{x_{m+1} x_n} = [\varphi_{x_n} , \varphi_{x_m} ] 
  \, ,  \qquad m,n = 1,2,  \ldots \; .   \label{pdCM_hier}
\ee
The circle closes by observing that this is ``pseudodual'' 
to the hierarchy (\ref{Ward_hier}) of Ward's chiral model in the following sense. 
(\ref{pdCM_hier}) is solved by 
\be
    \varphi_{x_n} = - J^{-1} \, J_{x_{n+1}}  \, , \qquad n=1,2,\ldots \, ,
         \label{varphi-J-system}
\ee
and the integrability condition of the latter system is the hierarchy (\ref{Ward_hier}). 
Rewriting (\ref{varphi-J-system}) as $J_{x_{n+1}} = - J \, \varphi_{x_n}$, the 
integrability condition is the hierarchy (\ref{pdCM_hier}). 
All this indeed connects the Ward model with the KP equation,  
but more closely with its matrix version, and not quite on a level which would 
allow a closer comparison of solutions. Note that the only nonlinearity that 
survives in the dispersionless limit is the commutator term, but this drops 
out in the ``projection'' to scalar KP. 
On the other hand, we established relations between \emph{hierarchies}, 
which somewhat ties their solution structure together.\footnote{We note, however, 
that e.g. the singular shock wave solutions of the dispersionless limit of 
the \emph{scalar KdV} equation have little in common with KdV solitons.}

In the Ward model, $J$ has values in $SU(m)$, thus $\varphi$ must 
have values in the Lie algebra $su(m)$, so has to be traceless and 
anti-Hermitian. Suitable conditions have to be imposed on $\Phi$ to achieve this. 
Via the dispersionless limit, methods of constructing exact 
solutions can be transfered from the (matrix) KP hierarchy 
to the \emph{pseudodual chiral model} (pdCM) hierarchy (\ref{pdCM_hier}). 
 From \cite{DMH07disp} we recall the following result. 
It determines in particular various classes of (multi-) lump solutions of the $su(m)$ pdCM 
hierarchy.

\begin{theorem}[]
\label{thm1}
Let $P,T$ be constant $N \times N$ matrices such that 
$T^\dagger = -T$ and $P^\dagger = T P T^{-1}$,  
and $V$ a constant $N \times m$ matrix. 
Suppose there is a constant solution $K$ of 
$[P,K] = -VV^\dagger T$ ($=Q$) such that 
$K^\dagger = T K T^{-1}$. 
Let $X$ be an $N \times N$ matrix solving $[X,P]=0$, 
$X^\dagger = T X T^{-1}$ and $X_{x_{n+1}} = X_{x_1} \, P^n$, $n=1,2,\ldots$. 
Then  $\varphi := -V^\dagger T (X - K)^{-1} V$ solves the 
$su(m)$ pdCM hierarchy. 
\end{theorem}

\vskip.1cm
\noindent
\textbf{Example~1.} Let $m=2$, $N=2$, and 
\be
    P = \left( \begin{array}{cc} p & 0 \\ 0 & p^\ast  \end{array} \right) \, ,
        \quad
    T = \left( \begin{array}{cc} 0 & -1 \\ 1 & 0  \end{array} \right) \, ,
        \quad
    X = \left( \begin{array}{cc} f & 0 \\ 0 & f^\ast  \end{array} \right) \, ,
        \quad 
    V = \left( \begin{array}{cc} a & b \\ c & d  \end{array} \right) \, , 
\ee
with complex parameters $a,b,c,d,p$ and a function $f$ (with complex conjugate 
$f^\ast$). Then $X_{x_{n+1}} = X_{x_1} \, P^n$ is satisfied if $f$ is an arbitrary 
holomorphic function of 
\be
   \omega := \sum_{n \geq 1} x_n \, p^{n-1} \; .   \label{omega}
\ee
Furthermore, $[P,K] = -VV^\dagger T$ 
has a solution iff $a c^\ast + b d^\ast =0$ and $\beta := 2\Im(p) \neq 0$ (where 
$\Im(p)$ denotes the imaginary part of $p$). Without restriction of generality 
we can set the diagonal part of $K$ to zero, since it can be absorbed by 
redefinition of $f$ in the formula for $\varphi$. 
We obtain the following components of $\varphi$, 
\be
   \varphi_{11} &=& - \varphi_{22} 
 = \frac{ i \, \beta}{D} \Big( |bc|^2-|ad|^2 + 2 \beta \, \Im(a^\ast c f) 
   \Big) \, , \nonumber \\
   \varphi_{12} &=& - \varphi_{21}^\ast 
 = \frac{\beta}{D} \Big( -2i \, (|c|^2+|d|^2) \, a^\ast b 
    + \beta \, (a^\ast d \, f - b c^\ast f^\ast) \Big) \, ,
          \label{1lump}
\ee
where $D := (|a|^2+|b|^2)(|c|^2+|d|^2) + \beta^2 \, |f(\omega)|^2 >0$ 
if $\det(V) \neq 0$. If $f$ is 
a non-constant polynomial in $\omega$, the solution is regular, rational and 
localized. It describes a simple lump if $f$ is linear in $\omega$. 
Otherwise it attains a more complicated shape (see \cite{DMH07disp} for 
some examples). 
\hfill $\square$
\vskip.1cm

Fixing the values of $x_4,x_5,\ldots$, we concentrate on the first pdCM 
hierarchy equation. 
In terms of the variables $x,y,t$ given by (\ref{coord_transf}), we then have
$\omega = \frac{1}{2}( t-x + 2 p y + p^2 (t+x) )$, subtracting a constant that 
can be absorbed by redefinition of the function $f$ in the solution in example~1. 
This solution becomes \emph{stationary}, i.e. $t$-independent, if $p= \pm i$. 
The conserved density 
\be
 \mathcal{E} := - \mathrm{tr}[(\varphi_t - \varphi_x)^2 + \varphi_y{}^2]/2 \; . 
                  \label{E}
\ee
of (\ref{pdCMv2}) is non-negative and will be used below to display the 
behaviour of some solutions. 
\vskip.1cm

More complicated solutions are obtained by superposition in the following sense. 
Given data $(X_1, P_1, T_1, V_1)$ and $(X_2, P_2, T_2, V_2)$ that determine 
solutions according to theorem~\ref{thm1}, we build  
\be
  P = \left(\begin{array}{cc} P_1 & 0 \\ 0 & P_2 \end{array}\right) \, , \qquad
    X = \left(\begin{array}{cc} X_1 & 0 \\ 0 & X_2
                       \end{array}\right)  \, , \qquad
  T = \left(\begin{array}{cc} T_1 & 0 \\ 0 & T_2 \end{array}\right) \, , \qquad
    V = \left(\begin{array}{c} V_1 \\V_2 \end{array}\right) \; . 
\ee
The diagonal blocks of the new big matrix $K$ will be $K_1$ and $K_2$. 
It only remains to solve 
\be
    P_1 K_{12} - K_{12} P_2 = - V_1 V_2^\dagger T_2 
\ee
for the upper off-diagonal block of 
$K$ and set $K_{21} = T_2^{-1} K_{12}^\dagger T_1$. In particular, one can 
superpose lump solutions as given in the preceding example.

\vskip.1cm
\noindent
\textbf{Example~2.} Superposition of two single lumps with $V_1=V_2=I_2$, 
the $2 \times 2$ unit matrix, yields
\be
   \varphi_{11} &=& -\varphi_{22} = -\frac{i}{\mathcal{D}} \Big( 
        \beta_2 |a h_1|^2 + \beta_1 |a h_2|^2 
      + 2 \beta_1 \beta_2 \, \Im(a^\ast h_1 h_2^\ast) + (\beta_1+\beta_2) |b|^4 
          \Big) \, , \qquad \nonumber \\
   \varphi_{12} &=& -\varphi_{21}^\ast
    =  \frac{1}{\mathcal{D}} \Big( 
       a |h_1|^2 \beta_2 h_2 + a^\ast \beta_1 h_1 |h_2|^2 
       + (b^\ast)^2 (a \beta_1 h_1 + a^\ast \beta_2 h_2) \Big) \, ,
\ee
where $\beta_i := 2 \, \Im(p_i)$, $a := p_1 - p_2^\ast$, $b := p_1 - p_2$, 
$h_1 := a \beta_1 f_1$, $h_2 := a^\ast \beta_2 f_2$ with arbitrary holomorphic 
functions $f_1(\omega_1)$ (where $\omega_1$ is (\ref{omega}) built with $p_1$), 
respectively $f_2(\omega_2)$, and 
$\mathcal{D} := (|b|^2 + |h_1|^2)(|b|^2 + |h_2|^2) + \beta_1 \beta_2 |h_1 - h_2|^2$. 
This solution is again regular if $p_1 \neq p_2$ \cite{DMH07disp}. 
 For $|f_1| \to \infty$ (resp. $|f_2| \to \infty$) we recover the single lump 
solution (\ref{1lump}) with $V=I_2$ and $f$ replaced by $f_2$ (resp. $f_1$).

\begin{figure}[t] 
\begin{center} 
\resizebox{14cm}{!}{
\includegraphics{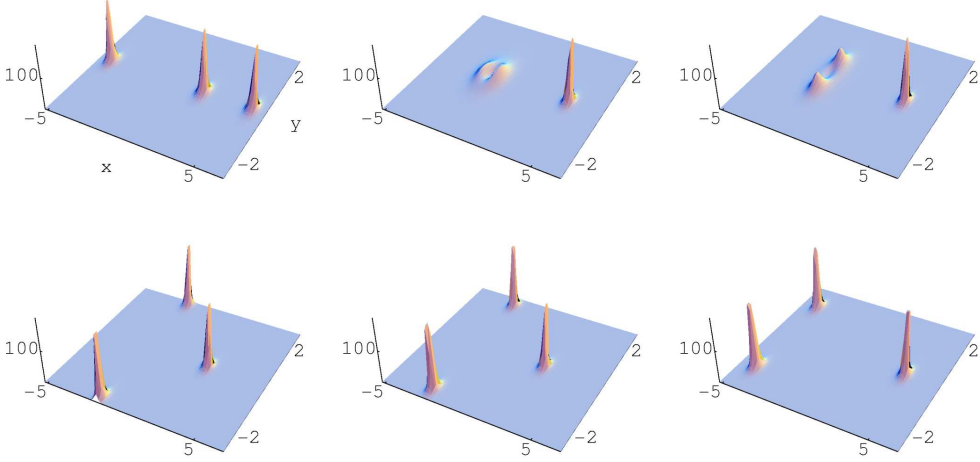}
}
\end{center} 
\caption{Plots of $\mathcal{E}$ at times $t=-90,-55,-53,0,30,80$ for the solution 
in example~2 with $p_1= -i (1-\epsilon)$ and $p_2=i (1+\epsilon)$ where $\epsilon = 1/20$, 
$f_1(\omega_1) = 4i \, \omega_1$, $f_2(\omega_2) = i \, \omega_2^2$. 
\label{fig1} }
\end{figure} 

\begin{figure}[t] 
\begin{center} 
\resizebox{16cm}{!}{
\includegraphics{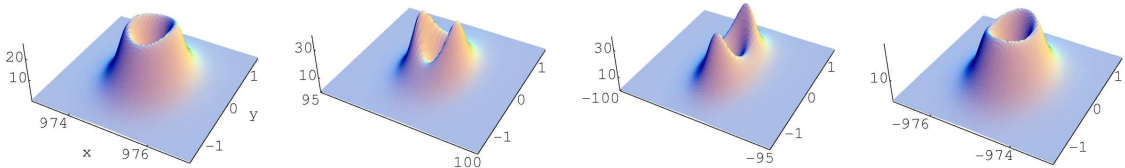}
}
\end{center} 
\caption{Origin and fate of the lump pair parts appearing in 
Fig.~1 (to the right in the first plot and to the left in the last). 
Plots of $\mathcal{E}$ at $t=-20000,-2000,2000,20000$. 
\label{fig2} }
\end{figure} 

Choosing $p_1 = i (1-\epsilon)$ and $p_2=i(1+\epsilon)$  
(or correspondingly with $i$ replaced by $-i$) with $0< \epsilon \ll 1$, 
and $f_1,f_2$ linear in $\omega_1$, respectively $\omega_2$, one observes 
scattering at right angle (cf. \cite{Ward95} for the analogous case in the Ward model). 

If $p_1 = -i (1-\epsilon)$ and $p_2=+i(1+\epsilon)$, 
one observes the following phenomenon: two lumps approach 
one another, meet, then separate in the orthogonal direction up to some maximal 
distance, reproach, merge again, and then separate again while moving in the 
original direction \cite{DMH07disp}.\footnote{See also \cite{LTG04} 
for an analogous phenomenon in case of KP-I lumps.}
In the limit $\epsilon \to 0$, $a$ vanishes and $\varphi$ becomes constant 
(assuming $f_1,f_2$ independent of $\epsilon$), so that $\mathcal{E}$ vanishes. 
For other choices of $f_1$ and $f_2$ 
more complex phenomena occur, including a kind of ``exchange process'' 
described in the following. Fig.~1 shows plots of $\mathcal{E}$ 
at successive times $t$ for the above solution with $f_1$ linear in 
$\omega_1$ and $f_2$ quadratic in $\omega_2$. The latter function then corresponds 
to a bowl-shaped lump (see the left of the plots in Fig.~2) which, at early times, 
moves to the left along the $x$-axis, deforming into the lump \emph{pair}, shown on the 
right hand side of the first plot in Fig.~1, under the increasing influence of the 
simple lump (corresponding to the linear function $f_1$) that moves to the right. 
When the latter meets the first partner of the lump pair, 
they merge, separate in $y$-direction to a maximal distance, move back 
toward each other and then continue moving as a lump pair (shown on the left 
hand side of the last plot in Fig.~1) into the negative $x$-direction. 
Meanwhile the remaining partner of the lump pair, that evolved from the original  
bowl-lump, retreats into the (positive) $x$-direction, with diminishing influence on 
the new lump pair, which then finally evolves into a bowl-shaped lump (see the right 
of the plots in Fig.~2). The smaller the value of $\epsilon$, the larger the range 
of the interaction. 
\hfill $\square$
\vskip.1cm

Other classes of solutions are obtained by taking for $P$ matrices of Jordan normal 
form, generalizing $T$ appropriately, and building superpositions in the 
aforementioned sense. Some examples in the $su(2)$ case have been worked out 
in \cite{DMH07disp}. This includes examples exhibiting (asymptotic) $\pi/n$ 
scattering of $n$-lump configurations. The pdCM (and also the Ward model) thus 
exhibits surprisingly complex lump interaction patterns, 
which are comparatively well accessible via the above theorem, though a kind 
of systematic classification is by far out of reach. 
\vskip.1cm

\noindent
\textbf{Acknowledgement.}
F M-H would like to thank the German Research Foundation for financial support 
to attend the workshop \emph{Algebra, Geometry, and Mathematical Physics} in 
G\"oteborg.

\end{document}